\documentclass[twocolumn,nopacs,floatfix,amsmath,nofootinbib,amssymb,preprintnumbers,floatfix]{revtex4}
\usepackage{amsmath}
\usepackage{longtable,lscape,booktabs}
\usepackage{txfonts}
\usepackage{overpic}
\usepackage{amssymb}
\usepackage{float}
\usepackage{placeins}
\usepackage{slashed}
\usepackage{color}
\usepackage{longtable,lscape,bm}
\usepackage{multirow}
\usepackage{graphicx,color,dcolumn,bm}
\usepackage[colorlinks,
            citecolor=blue,
            anchorcolor=red,
            menucolor=red,
            linkcolor=red,
            filecolor=red,
            runcolor=red,
            urlcolor=blue,
            frenchlinks=red]{hyperref}

\allowdisplaybreaks

\begin{document}

\title{Light-cone Distribution Amplitudes of Vector Mesons within the Self-consistent Light-front Quark Model}

\author{Xiao-Nan Li$^{1}$}\email{lixn@tlu.edu.cn}
\author{Shuai Xu$^{2}$} \email{xushuai@zknu.edu.cn (Corresponding author)}
\author{Qin Chang$^{3}$} \email{changqin@htu.edu.cn}

\affiliation{$^1$School of Electrical and Information Engineering, Tongling University, Tongling 244061, China\\
$^2$School of Physics and Telecommunications Engineering, Zhoukou Normal University, Zhoukou 466001, China\\
$^3$Institute of Particle and Nuclear Physics, Henan Normal University, Xinxiang 453007, China\\}

\begin{abstract}
In this paper, we investigate the twist-2 and twist-3 light-cone distribution amplitudes (LCDAs) of vector mesons within the self-consistent light-front quark model, and the $n$th Gegenbauer moment $a_n(\mu)$, $\xi$-moment $\langle \xi^n\rangle^{\parallel(\perp)}$ and transverse moment $\langle \mathbf{k}^n_\perp\rangle^{\parallel(\perp)}$ are also carried out. Adopting the parameter set fixed by confinements from the mesonic decay constants, we perform the numerical analysis and the results reveal several key insights: (i) The flavor symmetry breaking effects are more pronounced in the twist-3 LCDAs of vector mesons, which leads to the establishment of $a_1^\perp>a_1^\parallel$ and $\langle \xi^n\rangle^\perp>\langle \xi^n\rangle^\parallel$. This is consistent with previous findings in research for the LCDAs of pseudoscalar mesons. (ii) For vector mesons, the twist dependence decreases in the heavy quark limit which lead to $\phi_{2}^\parallel(x)\simeq\phi_{3}^\perp(x)$. For pseudoscalar and vector mesons composed of the same quark constituents, their LCDAs with the same twist exhibit similarity and gradually converge as the increasing of quark mass, i.e., $\phi_{2}^A(x)\simeq\phi_{2}^\parallel(x)$ and $\phi_{3}^P(x)\simeq\phi_{3}^\perp(x)$ in $m_q\rightarrow\infty$. In the heavy-quark limit within the self-consistent LFQM framework, we find an approximate correlated spin-twist independence pattern, with $\phi_2^A(x) \simeq \phi_2^{\parallel}(x) \approx\phi_3^P(x) \simeq \phi_3^{\perp}(x)$, resulting from the suppression of twist dependence and the approximate spin independence between pseudoscalar and vector mesons.

\end{abstract}

\maketitle

\section{introduction}

The hadronic LCDAs are typically defined as the matrix elements of nonlocal operators between the bound state and vacuum, which provide the important inputs in phenomenological researches of hard exclusive process via the factorization theorem~\cite{Lepage:1980fj,Brodsky:1997de,Efremov:1979qk}. The resulting factorization formulae of the exclusive amplitudes can be expressed in terms of heavy-to-light form factors, LCDAs, and perturbatively calculable short-distance coefficient functions, in analogy to the factorized expressions for the charmless hadronic $B$-meson decays~\cite{Huang:2024xii,Beneke:1999br}. In addition, the Gegenbauer moments are necessary inputs within the QCDF framework and play an indispensable role in calculating vertex corrections, spectator scattering and annihilation
contributions, especially exerting a substantial influence on annihilation contributions~\cite{Lu:2022kos}. The twist-2 LCDAs describe the distribution of valence quarks in terms of the longitudinal momentum fractions $x=p^+/P^+$ inside the bound state, by integrating the transverse momentum. The twist-3 and higher-twist LCDAs describe the distribution from transverse motion in the leading Fock state and the contribution from higher Fock states, such as quark-antiquark pair~\cite{Choi:2014kea}. It follows naturally that many nontrivial mesonic properties, such as decay constants and various form factors, are mutually influenced by and intertwined with the LCDAs. They also offer a valuable testing plant to understand the internal structure of hadrons and dynamic of QCD.

Theoretically, numerous approaches have been applied to compute these quantities, including QCD sum rules (QCDSR)~\cite{Chernyak:1983ej,Colangelo:2000dp,Bakulev:2005cp,Ball:1998sk,Yang:2007zt,Braguta:2008qe,Zhang:2025qmg,Zeng:2025rfe,Wang:2025sic,Han:2013zg,Zhong:2021epq,Khodjamirian:2004ga,Khodjamirian:2020hob}, lattice QCD (LQCD)~\cite{CP-PACS:2001vqx,Braun:2006dg,LatticeParton:2024vck,LatticeParton:2024zko,LatticeParton:2022zqc,Ding:2024saz,Baker:2024zcd,Blossier:2024wyx,Zhang:2020gaj,Cloet:2024vbv}, perturbative QCD (PQCD)~\cite{Chai:2025xuz,Cheng:2019ruz,Cheng:2020vwr}, Dyson--Schwinger equations (DSE)~\cite{Maris:1997tm,Roberts:1994dr,Chang:2013pq,Chang:2013epa,Shi:2015esa,Roberts:2021nhw,Xu:2025hjf}, chiral quark models~\cite{Petrov:1998kg,Nam:2006au,Son:2024uet,Broniowski:2007si}, Nambu-Jona-Lasinio models~\cite{RuizArriola:2002bp,Praszalowicz:2001wy,Noguera:2015iia,Courtoy:2019cxq}, and the light-front quark model (LFQM)~\cite{Jaus:1999zv,Choi:2007yu,Hwang:2008qi,Ji:1992yf,Terentev:1976jk,Chen:2021ywv,Chang:2020wvs}. Leveraging the advantages of manifest Lorentz invariance and conceptual simplicity, the LFQM has been quite successfully used in analyzing various hadronic form factors, decay constants, LCDAs, etc~\cite{Xu:2025aow,Xu:2025ntz,Li:2026wmb,Chang:2019obq,Chang:2019mmh}.

The LFQM is based on the light-front (LF) dynamics which carries the maximum number (seven) of the kinematic (or interaction-independent) generators and thus less effort in dynamics is necessary in order to get the QCD solutions that reflect the full Poincar\'e symmetries~\cite{Dirac:1949cp,Brodsky:1998hn,Keister:1991sb,Coester:1992cg,Szczepaniak:1995vn,Osborn:1972dy,Susskind:1967rg,Bardakci:1968zqb,Chang:1968bh,Brodsky:1973kb,Bjorken:1970ah,Neville:1971uc}. The basic ingredient in LFQM is the relativistic hadron wave function which contains all information of a bound state in principle and generalizes the LCDAs by integrating the transverse momentum. The hadronic quantities are represented by the overlap of wave functions and can be derived naturally. In the LFQM, hadrons are composed of valence quarks and the equation of motion for the bound state is a relativistic Schr\"{o}dinger equation with an effective confining potential, e.g., $\mathcal{H}_{LF}|\Psi\rangle=M|\Psi\rangle$, where the LF Hamiltonian of the bound state $\mathcal{H}_{LF}=H_0+V_{q\bar{q}}$ ~\cite{Brodsky:2014yha} and $|\Psi\rangle$ contains radial and spin components. Generally, the former $\psi(x,\mathbf{k}_\perp)$ is obtained by solving the LF equation, while the fully relativistic treatment of quark spins and the center-of-mass motion can be carried out using the Melosh rotation~\cite{Melosh:1974cu}. Rather than calculating these wave functions from a phenomenological potential, one often starts with an empirical trial wave function. For instance, the popular phenomenological Gaussian-type wave functions have been suggested in the literature.

Our previous studies have mainly focused on pseudoscalar mesons and their LCDAs~\cite{Li:2026wmb}. Although some theoretical ingredients can be formally generalized to vector mesons, the extension is highly nontrivial because vector mesons involve richer spin structures, polarization dependence, and distinct twist decompositions. In particular, it remains unclear whether the spin-independent behavior observed in heavy pseudoscalar systems can persist in vector mesons and how it correlates with the twist structure of the LCDAs. A systematic investigation of vector meson LCDAs therefore provides an important testing ground for understanding the interplay between spin, twist, and heavy-quark dynamics within the self-consistent LFQM framework. Phenomenologically, several flavor anomalies have been observed, such as the discrepancy between the angular observable $P_5^\prime(B \to K^{*0}\mu^+\mu^-)$ measured by the LHCb collaboration and the Standard Model (SM) predictions.
In order to make precise theoretical predictions for observables in $B \to K^{*0}\mu^+\mu^-$ decays, high-precision calculations of the $B \to K^{*0}$ form factors are of paramount importance~\cite{Gao:2024vql}. The $R(D_{(s)}^{(*)})$ anomalies, which reflect lepton flavor universality violating effects, have long been considered compelling indications of ``New Physics'' beyond the SM~\cite{Cui:2023jiw,Shen:2021yhe,Gao:2019lta}. From a theoretical perspective, the LCDAs of vector mesons still exert a non-negligible influence on this semileptonic process.

In this work, we utilize the LFQM framework to study the LCDAs of vector mesons. Meanwhile, we present three key physical moments: the Gegenbauer moment $a_n(\mu)$, which describes the degree of deviation of the LCDAs from their asymptotic form; the $\xi$-moment $\langle \xi^n \rangle$, which characterizes the longitudinal momentum distribution of the quarks; and the transverse moment $\langle \mathbf{k}_\perp^n \rangle$, which reflects the transverse scale of the bound state. These results provide important information for a deeper understanding of the internal structure of vector mesons.

The paper is organized as follows: Section~\ref{sec:2} provides a brief review of the theoretical formalism for the LFQM and the LCDAs. Numerical results and discussion are presented in Section~\ref{sec:3}. Finally, conclusions are given in Section~\ref{sec:4}.

\section{Preliminary}\label{sec:2}

In the LFQM, the Fock state is treated as in a noninteraction $q\bar{q}$ representation and the interactions are encoded in the LF wave function $\Psi_{h\bar{h}}(x,\mathbf{k}_\perp)$. Specifically, the expansion for a meson is given by
\begin{align}
|M(P)\rangle &= \int \{d^3p_q\}\{d^3p_{\bar{q}}\}2(2\pi)^3\delta^3(P-p_q-p_{\bar{q}}) \nonumber \\ &\times\sum_{h,\bar{h}}\Psi_{h\bar{h}}(x,\mathbf{k}_\perp)|q(p_q,h)\bar{q}(p_{\bar{q}},\bar{h})\rangle,
\label{eq:1}
\end{align}
where $p_q (p_{\bar{q}})$ and $h (\bar{h})$ is momentum and helicity of the constituent quark (antiquark), respectively. The momentum assignments in terms of LF variable $(x,\mathbf{k}_\perp)$ for constituents are as follows:
\begin{eqnarray}
p_q=(xP^+,\frac{m_q^2}{xP^+},\mathbf{k}_\perp),~~~~~~~ p_{\bar{q}}=(\bar{x}P^+,\frac{m_{\bar{q}}^2}{\bar{x}P^+},-\mathbf{k}_\perp),
\label{eq:2}
\end{eqnarray}
in which $\bar{x}=1-x$ and $P=p_q+p_{\bar{q}}$ is inherently satisfied. The LF wave function $\Psi_{h\bar{h}}(x,\mathbf{k}_\perp)$ is generally defined as
\begin{eqnarray}
\Psi_{h\bar{h}}(x,\mathbf{k}_\perp) = \psi(x,\mathbf{k}_\perp)S_{h\bar{h}}(x,\mathbf{k}_\perp),
\label{eq:3}
\end{eqnarray}
where $\psi(x,\mathbf{k}_\perp)$ is the radial wave function and $S_{h\bar{h}}(x,\mathbf{k}_\perp)$ corresponds to the spin wave function.

For vector mesons, one kind of 1S state radial wave function $\psi(x,\mathbf{k}_\perp)$ has been suggested in previous works~\cite{Xu:2025aow,Xu:2025ntz} and the well-validated Gaussian-type wave function is taken in this work
\begin{eqnarray}
\psi(x,{\bf{k}}_{\perp})= \frac{4\pi^{3/4}}{\beta^{3/2}}\sqrt{\frac{\partial k_z}{\partial x}} \mathrm{exp}(-\frac{{\bf{k}}_{\perp}^2+k_z^2}{2\beta^2}).
\label{eq:4}
\end{eqnarray}
Here, the Gaussian-type wave function should be regarded as a phenomenological ansatz rather than a first-principles QCD solution,
where the quark mass and $\beta$ are the only two inputs in the LFQM. With the confinements from mesonic decay constants, the parameter set could be well fixed and the details are presented in literature~\cite{Xu:2025aow,Xu:2025ntz}. The self-consistent LFQM is based on the Bakamjian--Thomas (BT) construction~\cite{Bakamjian:1953kh,Chung:1988my,Polyzou:2010zz}, in which a meson state is described by an on-shell quark--antiquark pair and interactions are incorporated through the mass operator $M = M_0 + V$, where $M_0$ is the internal invariant mass constructed from the light-front variables $(x, k_\perp)$. It is defined as $M_0=\sqrt{\frac{m_{q}^2+{\bf{k}}_{\perp}^2}{x}+\frac{m_{\bar{q}}^2+{\bf{k}}_{\perp}^2}{\bar{x}}}$. The invariant mass $M_0$ plays a crucial role in the analysis of covariance and self-consistency between the standard and covariant formulations of LFQM~\cite{Chang:2019obq,Chang:2019mmh}. Under the variable transformation $\{x,{\bf{k}}_{\perp}\}\rightarrow \vec{k}\equiv\{{\bf{k}}_{\perp},k_z\}$, the longitudinal momentum is defined as $k_z=(x-\frac{1}{2})M_0+\frac{m_q^2-m_{\bar{q}}^2}{2M_0}$, with the corresponding Jacobian $\frac{\partial k_z}{\partial x}= \frac{M_0}{4x\bar{x}}\Large\{1-\big[\frac{(m_q-m_{\bar{q}})^2}{M_0^2}\big]^2\Large\}$. At the meson-quark vertex, light-front energy conservation requires $P^- = p_q^- + p_{\bar q}^- $. This condition implies that the external meson mass entering the integrand requires consistency with invariant mass $M_0(x, k_\perp)$. Otherwise, the kinematical relation $(M^2 + P_\perp^2)/P^+=(m_q^2 + p_{q\perp}^2)/p_q^++(m_{\bar q}^2 + p_{\bar q\perp}^2)/{p_{\bar q}^+}$ would be violated. Therefore, the replacement $M \to M_0$ follows from enforcing light-front energy conservation within the BT framework, and is required for maintaining covariance and self-consistency.


For LF spin-orbit wave function, $S_{h\bar{h}}(x,\mathbf{k}_\perp)$ is obtained by the interaction independent Melosh transformation from the traditional spin-orbit wave function and the covariant form is given by
\begin{align}
S_{h\bar{h}}(x,\mathbf{k}_\perp)&= \frac{\bar{u}(p_q,h)\Gamma \nu(p_{\bar{q}},h)}{\sqrt{2}\sqrt{M_0^2-(m_q-m_{\bar{q}})^2}},\nonumber\\
\Gamma&=-\slashed{\epsilon}+\frac{\epsilon\cdot(p_q-p_{\bar{q}})}{M_0+m_q+m_{\bar{q}}},
\label{eq:5}
\end{align}
where $u$, $v$ are Dirac spinors and $\sum_{h,\bar{h}}S^\dagger S=1$.

The quark field could be expanded in terms of creation and annihilation operator as
\begin{align}
q(x)&=\int\frac{dp_q^+}{\sqrt{2p_q^+}}\frac{d^2\mathbf{p}_{q\perp}}{(2\pi)^3}\sum_h[b_h(p_q^+,\mathbf{p}_{q\perp})u_h(p_q^+,\mathbf{p}_{q\perp})e^{-ip_q\cdot x}\nonumber\\ &+d_h^\dagger(p_q^+,\mathbf{p}_{q\perp})\nu_h(p_q^+,\mathbf{p}_{q\perp})e^{ip_q\cdot x}]
\label{eq:6}
\end{align}
and the expression for $\bar{q}(x)$ can be obtained by taking the conjugate of Eq.~(\ref{eq:6}).
With the expressions of meson and quark (antiquark) field in Eqs.~(\ref{eq:1}) and (\ref{eq:6}), the matrix elements for $\langle 0|\bar{q}(z)\Gamma q(-z)|M(P)\rangle$ are derived as
\begin{eqnarray}
\mathcal{M}_{\Gamma}=\sqrt{N_c}\sum_{h,\bar{h}}\int\frac{dp_q^+}{\sqrt{2p_q^+}}\frac{d^2\mathbf{p}_\perp}{(2\pi)^3}
\Psi_{h,\bar{h}}(x,\mathbf{k}_\perp)\bar{\nu}_{\bar{h}}\Gamma u_h e^{i(2x-1)P\cdot z}.
\label{eq:7}
\end{eqnarray}

The LCDAs of a vector mesons are defined in terms of the following matrix elements
\begin{align}
\langle 0|\bar{q}(0)\gamma^\mu q(z)|V(P,l)\rangle &= f_V M\int_0^1 dx~e^{-i x P\cdot z}\{P^\mu\frac{\epsilon_l\cdot z}{P\cdot z}\phi_{2}^\parallel(x,\mu)\nonumber\\&+(\epsilon_l^\mu-P^\mu\frac{\epsilon_l\cdot z}{P\cdot z})\phi_{3}^\perp(x,\mu)+...\},
\label{eq:8}
\end{align}
where $\phi_{2}^\parallel(x,\mu)$ and $\phi_{3}^\perp(x,\mu)$ correspond to the twist-2 and twist-3 LCDAs. The $...$ denote the higher twist LCDAs which will be neglected below, and then the twist-2 and twist-3 LCDAs can be derived from Eq.~(\ref{eq:8}) as
\begin{eqnarray}
\langle 0|\bar{q}(0)\gamma^+ q(z^-)|V(P,l)\rangle = f_V M \epsilon_0^+\int_0^1 dx~e^{-i x P\cdot z}\phi_{2}^\parallel(x,\mu),
\label{eq:9}
\end{eqnarray}
for $\phi_{2}^\parallel(x,\mu)$ by taking the plus component $\mu=+$ of the current and the longitudinal polarization $l=0$ and
\begin{eqnarray}
\langle 0|\bar{q}(0)\gamma^\perp q(z^-)|V(P,l)\rangle = f_V M \epsilon_+^\perp\int_0^1 dx~e^{-i x P\cdot z}\phi_{3}^\perp(x,\mu),
\label{eq:10}
\end{eqnarray}
for $\phi_{3}^\perp(x,\mu)$ by taking the perpendicular component $\mu=\perp$ of the current and the transverse polarization $l=+$, respectively. The integration variable $x$ is the longitudinal momentum fraction of the quark and $\xi=x-\bar{x}=2x-1$ depicts longitudinal separation. By taking the Fourier transform for Eq.~(\ref{eq:9}) with the redefined variable $z^{\mu}=\tau\eta^{\mu}$ where the lightlike vector $\eta=(1,0,0,-1)$, we can obtain
\begin{eqnarray}
\int_{-\infty}^\infty d\tau \langle 0|\bar{q}(0)\gamma^+ q(-\tau\eta)|V(P,l)\rangle e^{-i\xi'\tau P\cdot \eta}=\nonumber\\f_V M \epsilon_0^+\int_{-\infty}^\infty d\tau \int_0^1 dx~\phi_{2}^\parallel(x,\mu)e^{i(\xi-\xi')\tau P\cdot \eta},
\label{eq:11}
\end{eqnarray}
in which $\xi'=2x'-1$, where dummy variable $x'$ corresponds to the conjugate variables for $\tau$. Substituting Eq.~(\ref{eq:7}) into the left-hand side of Eq.~(\ref{eq:11}), we can directly extract $\phi_{2}^\parallel(x,\mu)$ as
\begin{align}
\phi_{2}^\parallel(x,\mu)&=\frac{1}{f_VM \epsilon_0^+}\sqrt{N_c}\nonumber\\&\times\sum_{h,\bar{h}}\int\frac{dp_q^+}{\sqrt{2p_q^+}}\frac{d^2\mathbf{p}_\perp}{(2\pi)^3}
\Psi_{h,\bar{h}}(x,\mathbf{k}_\perp)\bar{\nu}_{\bar{h}}\gamma^+ u_h.
\label{eq:12}
\end{align}
Applying the LF wave function and performing the spinor contraction with $\bar{u}u=\slashed{p}$, we simplify the right-hand side of Eq.~(\ref{eq:12}). The calculation of trace term is straightforward and the result for twist-2 LCDA is carried out:
\begin{align}
\phi_{2}^\parallel(x,\mu) &= \frac{\sqrt{2N_c}}{f_V}\int_0^{\mu^2} \frac{d^2\mathbf{k}^2_\perp}{8\pi^3}\frac{\psi(x,{\bf{k}}_{\perp})}{\sqrt{(\bar{x}m_q+xm_{\bar{q}})^2
+\mathbf{k}^2_\perp}}\nonumber\\&\times[(\bar{x}m_q+xm_{\bar{q}})+\frac{2\mathbf{k}^2_\perp}{M_0+m_q+m_{\bar{q}}}].
\label{eq:13}
\end{align}
Adopting the same procedure, the formula of twist-3 LCDA is derived as
\begin{align}
\phi_{3}^\perp(x,\mu) &= \frac{\sqrt{2N_c}}{f_V}\int_0^{\mu^2} \frac{d^2\mathbf{k}^2_\perp}{8\pi^3}\frac{\psi(x,{\bf{k}}_{\perp})}{\sqrt{(\bar{x}m_q+xm_{\bar{q}})^2
+\mathbf{k}^2_\perp}}\nonumber\\&\times[\frac{(\bar{x}m_q+xm_{\bar{q}})^2+\mathbf{k}^2_\perp}{2x\bar{x}M_0}-\frac{\mathbf{k}^2_\perp}{M_0+m_q+m_{\bar{q}}}].
\label{eq:14}
\end{align}
  In the view of QCD, the scale $\mu$ separates nonperturbative and perturbative regimes, which is introduced by a cutoff via $|\mathbf{k}_\perp|\leq\mu$ in transverse integration in the LFQM. Following the earlier LFQM analysis~\cite{Choi:2013mda,Choi:2014ifm,Choi:2017uos,Arifi:2022pal}, one observes that $|\mathbf{k}_\perp| \to \infty$ corresponds to the ultraviolet (UV) cutoff $\mathbf{k}_\perp^{\max}$ or typical energy scales $\mu$ around $1~\mathrm{GeV}$ for light mesons, $2~\mathrm{GeV}$ for ($D^{*}_{(s)}$,$B^{*}_{(s)}$) mesons, and $3~\mathrm{GeV}$ for $B_c^{*}$ meson.
Although moderate variations of $\mu$ may quantitatively affect some observables, the main qualitative conclusions of this work remain stable under reasonable changes of the cutoff scale. Without loss of generality, we adopt the same prescription in our numerical analysis. These LCDAs are usually expanded in terms of the Gegenbauer polynomials $C_n^{3/2}$ as follows:
\begin{eqnarray}
\phi_{2(3)}^{\parallel(\perp)}(x,\mu)&=&\phi^{\parallel(\perp)}_{as}(x)\sum_0^\infty a_n^{\parallel(\perp)}(\mu)C_n^{3/2}(2x-1),
\label{eq:15}
\end{eqnarray}
where the asymptotic LCDA $\phi^\parallel_{as}(x)=6x\bar{x}$ and $\phi^\perp_{as}(x)=\frac{3}{4}[1+(2x-1)^2]$ in the high energy limit. The explicit formula for $a_n(\mu)$ is given by
\begin{eqnarray}
a_n^{\parallel(\perp)}(\mu)&=&\frac{4n+6}{3n^2+9n+6}\int_0^1 dx C_n^{3/2}(2x-1)\phi^{\parallel(\perp)}_{2(3)}(x,\mu).
\label{eq:16}
\end{eqnarray}
We can obtain additional information about bound state's characteristics such as the $\xi$-moment
\begin{eqnarray}
\langle\xi^n\rangle^{\parallel(\perp)} = \int_0^1 dx~\xi^n~\phi^{\parallel(\perp)}_{2(3)}(x,\mu),
\label{eq:17}
\end{eqnarray}
where $\xi$ represents the longitudinal discrepancy between constituent quarks in the bound state. Similarly, the nonperturbative quantity transverse moment is also obtained by
\begin{eqnarray}
\langle\mathbf{k}^n_\perp\rangle^{\parallel(\perp)} = \int_0^1 dx~\mathbf{k}^n_\perp~\phi^{\parallel(\perp)}_{2(3)}(x,\mu),
\label{eq:18}
\end{eqnarray}
which reflects the information about the transverse size of the bound state.

\section{Numerical results}\label{sec:3}

\begin{figure*}[!htbp]
\centering
\includegraphics[width=6cm,height=5cm]{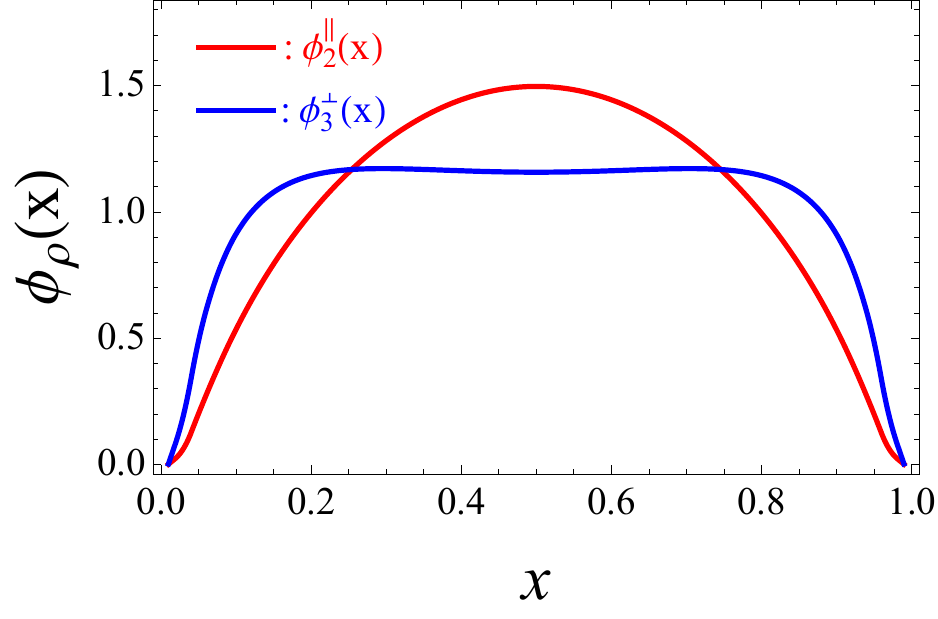}~~~~~~~~~~~
\includegraphics[width=6cm,height=5cm]{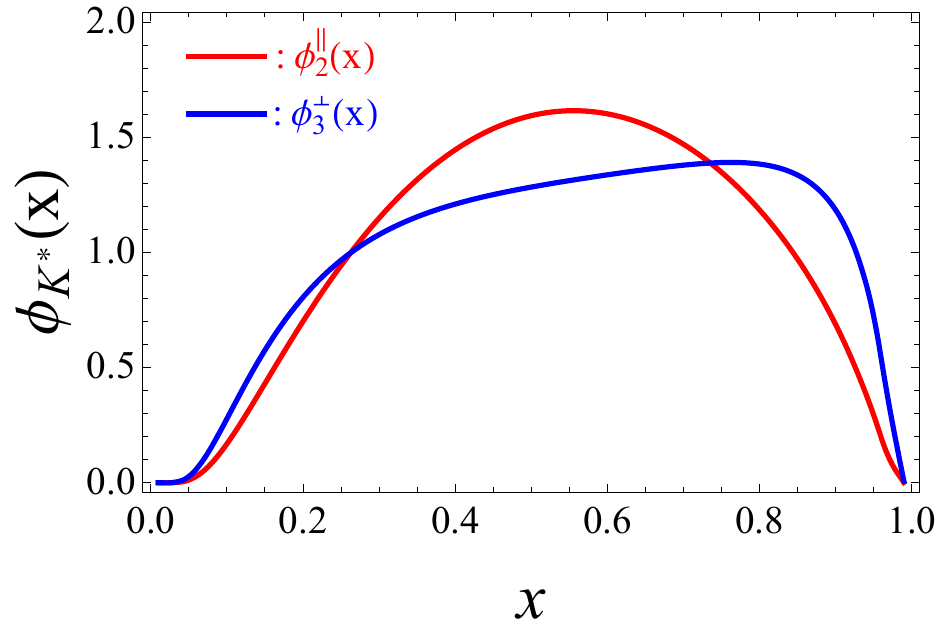}
\caption{The twist-2 $\phi_2^\parallel(x)$ (the red line) and twist-3 $\phi_3^\perp(x)$ (the blue line) LCDAs of $\rho$ and $K^*$.}
\label{1}
\end{figure*}

\begin{figure*}[!htbp]
\centering
\includegraphics[width=6cm,height=5cm]{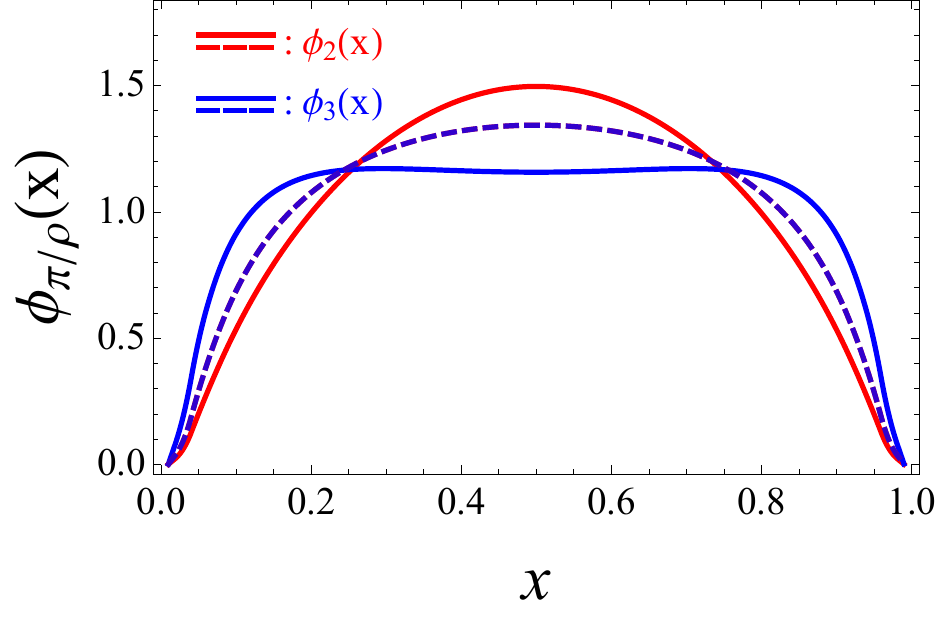}~~~~~~~~~~~
\includegraphics[width=6cm,height=5cm]{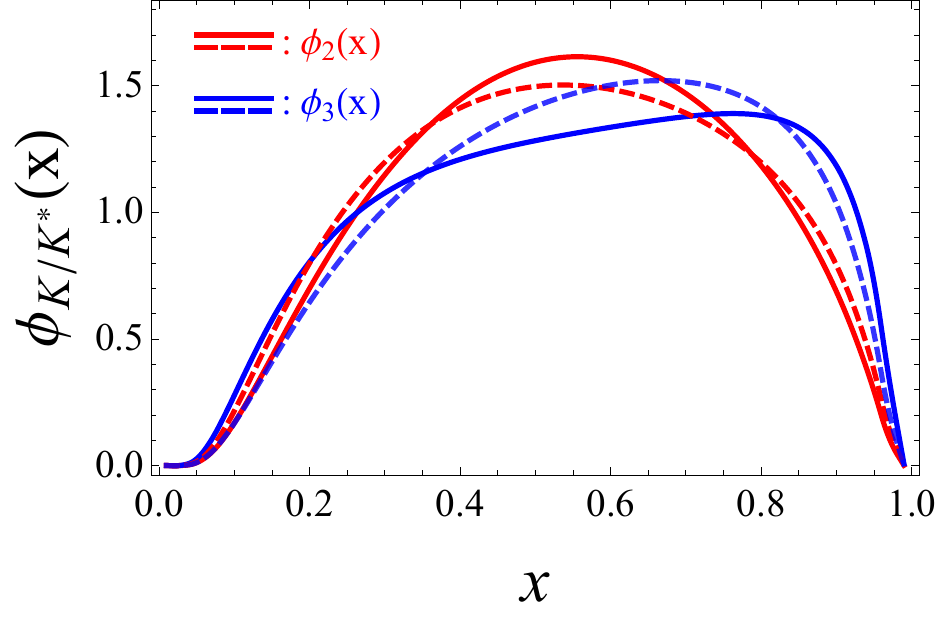}
\caption{The twist-2 $\phi_2(x)$ ($\phi_2^A(x)$, $\phi_2^\parallel(x)$) and twist-3 $\phi_3(x)$ ($\phi_3^P(x)$, $\phi_3^\perp(x)$) LCDAs of $\pi$/$\rho$ and $K$/$K^*$ mesons, in which the dashed and solid lines are the LCDAs of pseudoscalar and vector mesons, respectively.}
\label{2}
\end{figure*}

\begin{figure*}[!htbp]
\centering
\includegraphics[width=5cm,height=4cm]{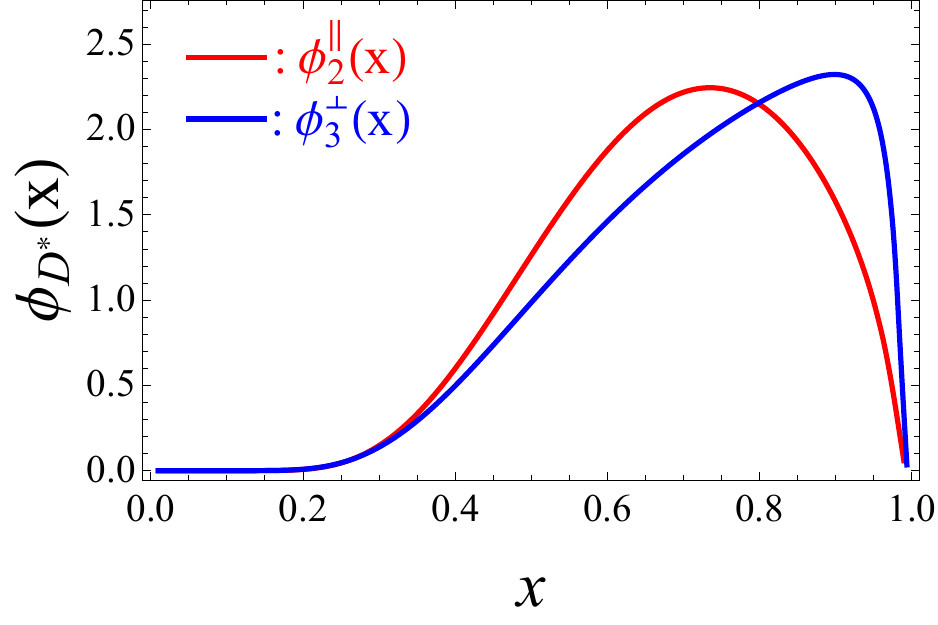}~~~~~~~~
\includegraphics[width=5cm,height=4cm]{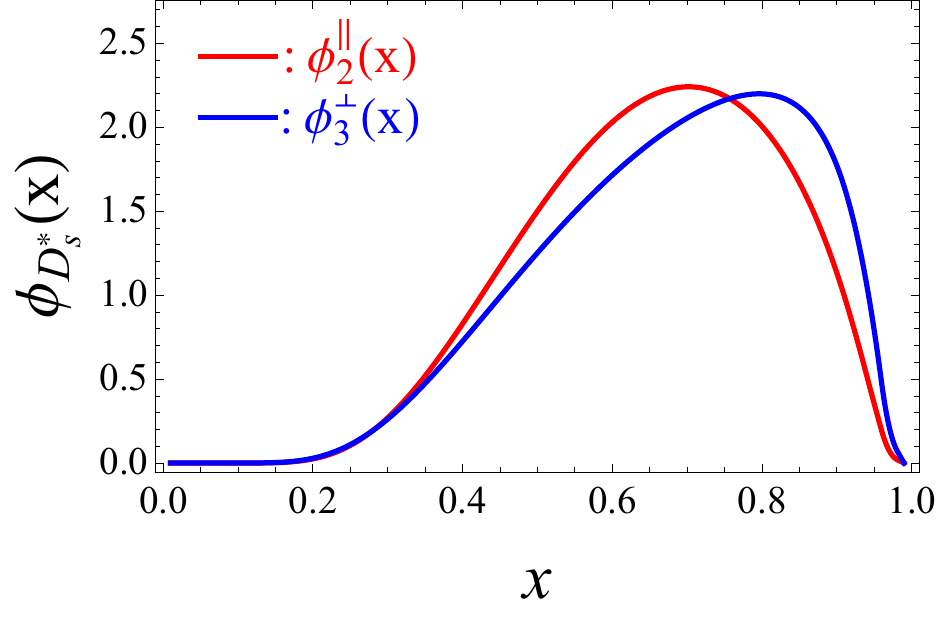}\\
\includegraphics[width=5cm,height=4cm]{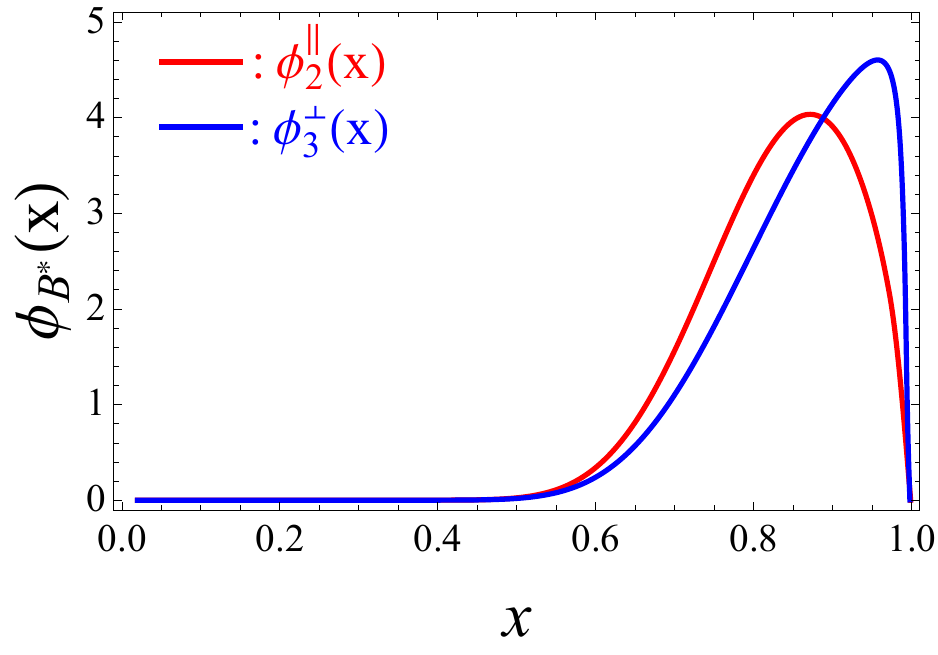}~~
\includegraphics[width=5cm,height=4cm]{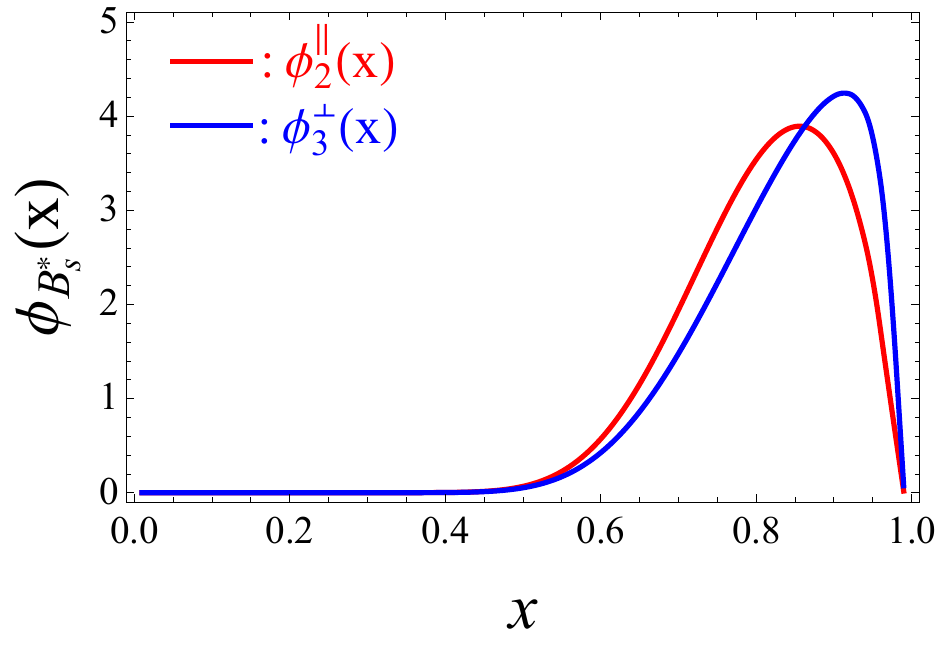}~~
\includegraphics[width=5cm,height=4cm]{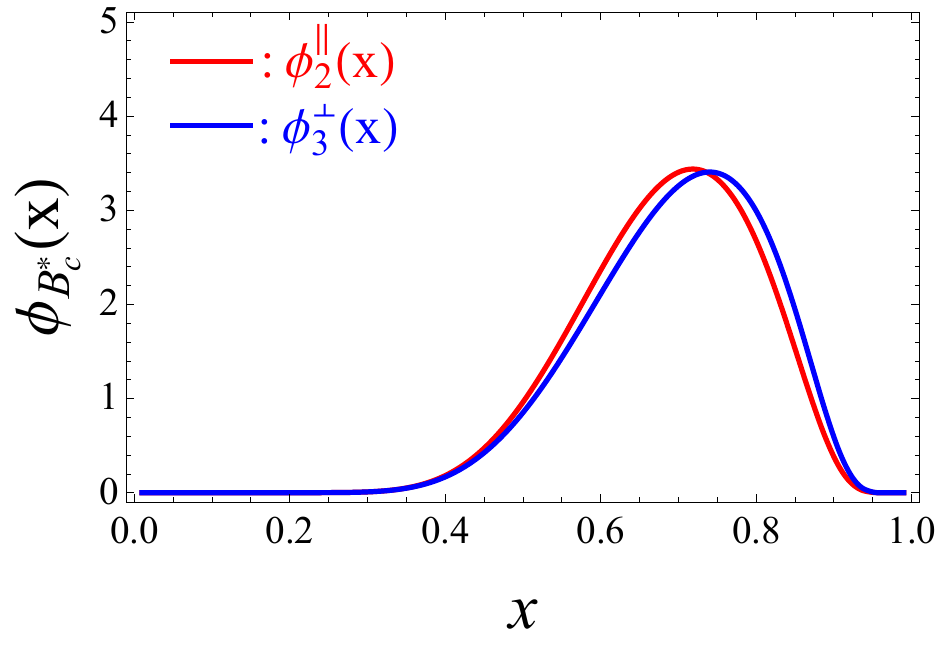}
\caption{Same as in FIG.~\ref{1} but with the $D^*_{(s)}$ and $B^*_{(s/c)}$ mesons.}
\label{3}
\end{figure*}

\begin{figure*}[!htbp]
\centering
\includegraphics[width=5cm,height=4cm]{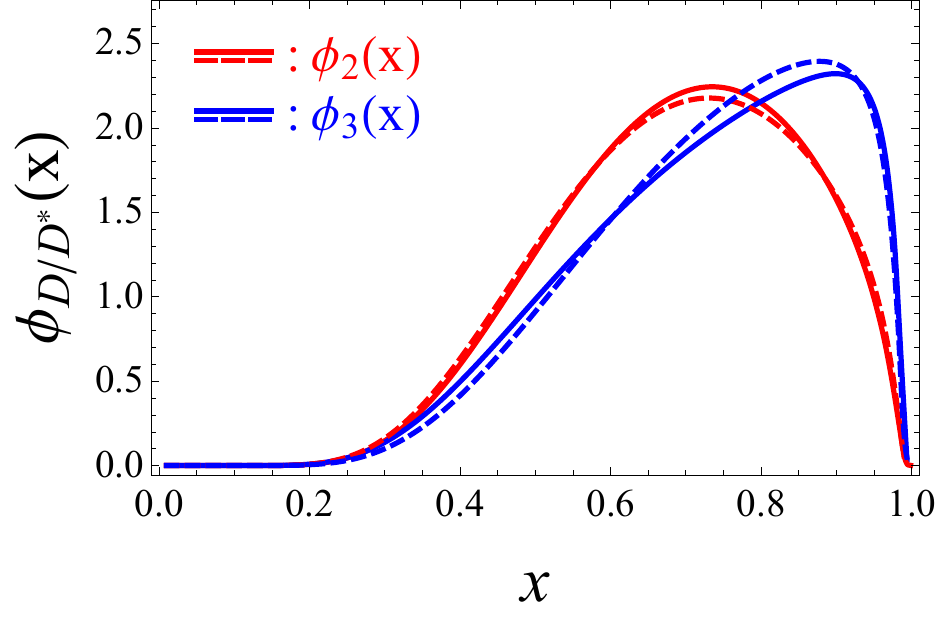}~~~~~~~~
\includegraphics[width=5cm,height=4cm]{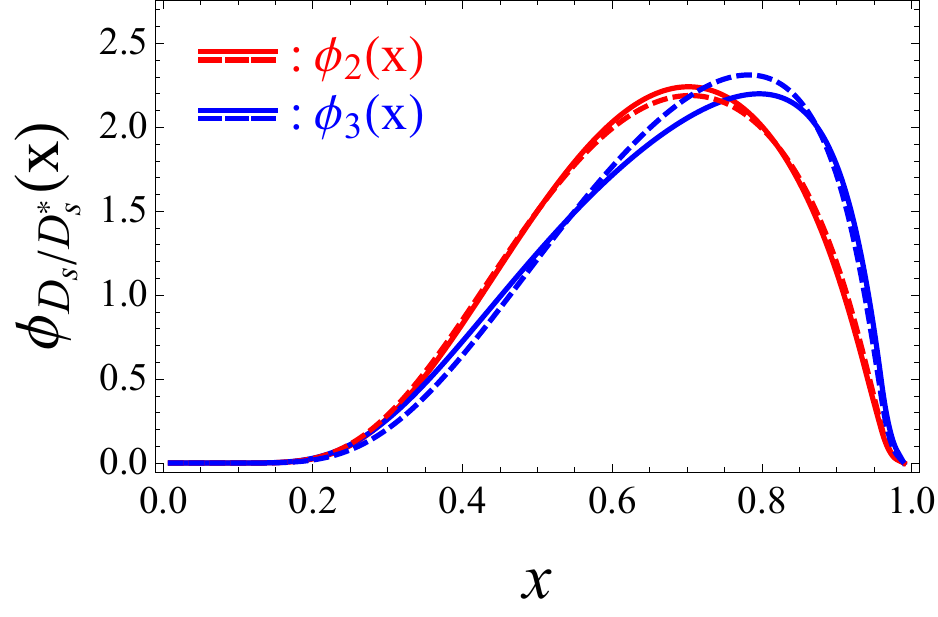}\\
\includegraphics[width=5cm,height=4cm]{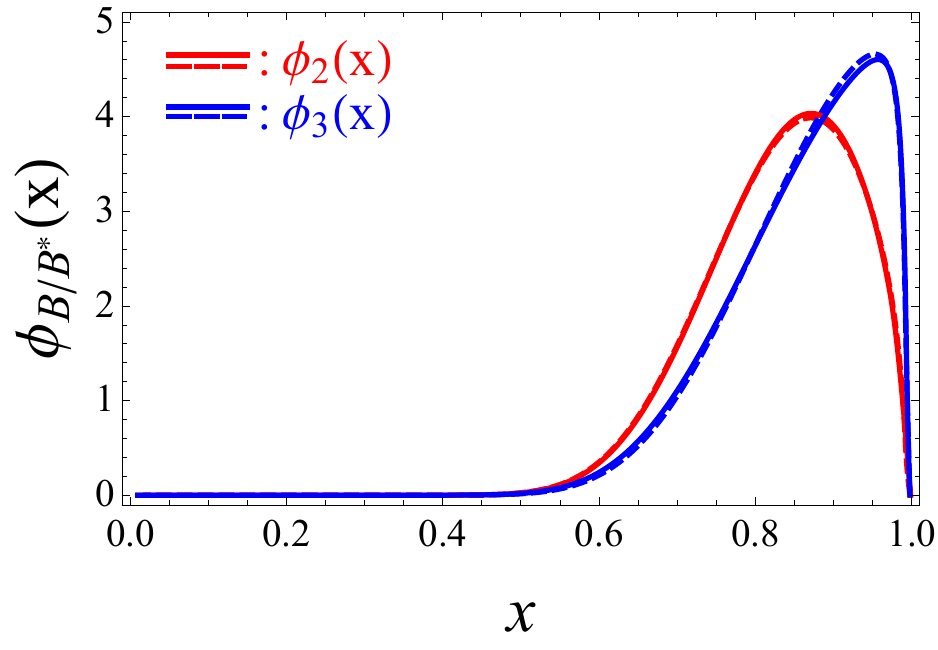}~~
\includegraphics[width=5cm,height=4cm]{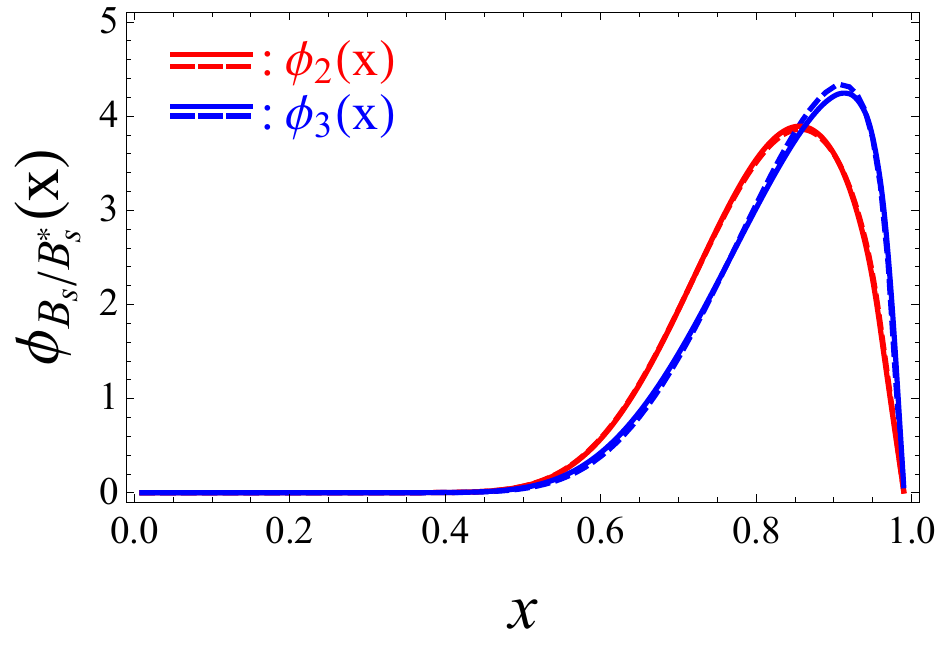}~~
\includegraphics[width=5cm,height=4cm]{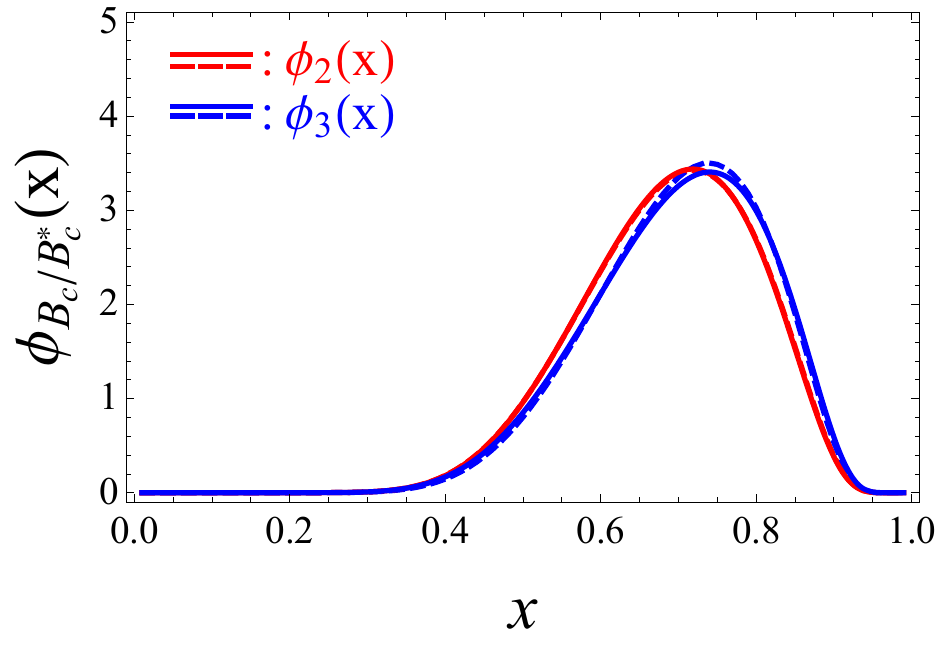}
\caption{Same as in FIG.~\ref{2} but with the $D_{(s)}$/$D^*_{(s)}$ and $B_{(s/c)}$/$B^*_{(s/c)}$ mesons.}
\label{4}
\end{figure*}

The model parameters for the constituent quark masses and Gaussian parameter $\beta$ are listed in Table~\ref{tab:1}. They are mainly determined by fitting the mesonic decay constants reported by the Particle Data Group~\cite{ParticleDataGroup:2022pth}. The LCDAs of vector mesons are presented in Figs.~\ref{1}--\ref{4}, and several remarks are summarized below:
\begin{table*}[!htbp]
\caption{\label{tab:1}Quark masses $m$ and Gaussian parameters $\beta$ (in GeV).}
\centering
\footnotesize
\begin{tabular*}{\textwidth}{@{\extracolsep{\fill}}ccccccccccc}
\hline\hline
$m_q$ & $m_s$ & $m_c$ & $m_b$
& $\beta_{qq}$ & $\beta_{qs}$ & $\beta_{qc}$ & $\beta_{cs}$
& $\beta_{qb}$ & $\beta_{bs}$ & $\beta_{bc}$ \\
\hline
$0.25^{+0.01}_{-0.01}$ &
$0.50^{+0.02}_{-0.02}$ &
$1.80^{+0.09}_{-0.09}$ &
$5.10^{+0.25}_{-0.25}$ &
$0.321^{+0.016}_{-0.016}$ &
$0.352^{+0.017}_{-0.017}$ &
$0.465^{+0.023}_{-0.023}$ &
$0.522^{+0.026}_{-0.026}$ &
$0.535^{+0.026}_{-0.026}$ &
$0.594^{+0.003}_{-0.003}$ &
$0.883^{+0.044}_{-0.044}$ \\
\hline
\end{tabular*}

\end{table*}
\begin{figure*}[!htbp]
\centering
\includegraphics[width=6cm,height=5cm]{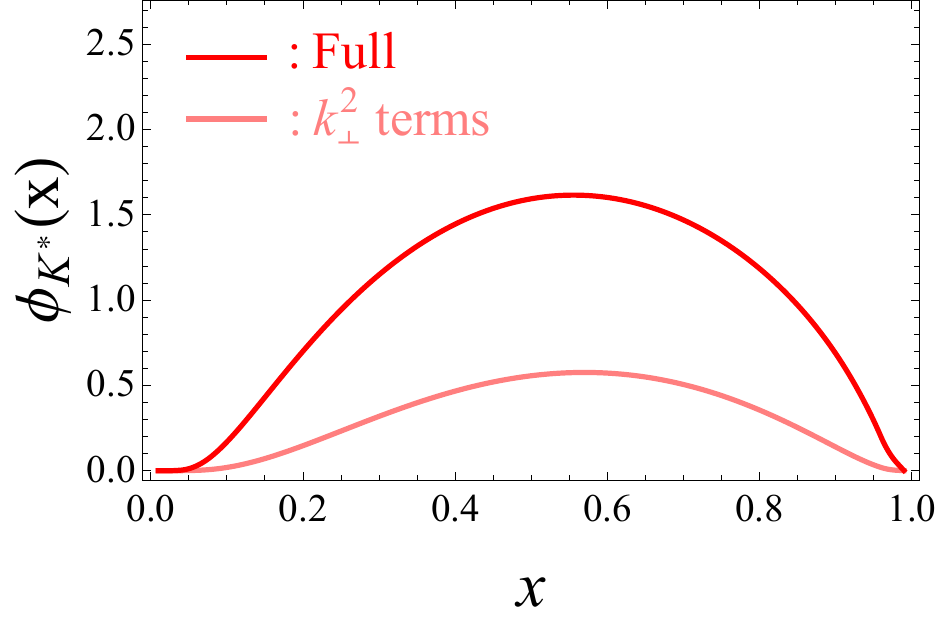}~~~~~~
\includegraphics[width=6cm,height=5cm]{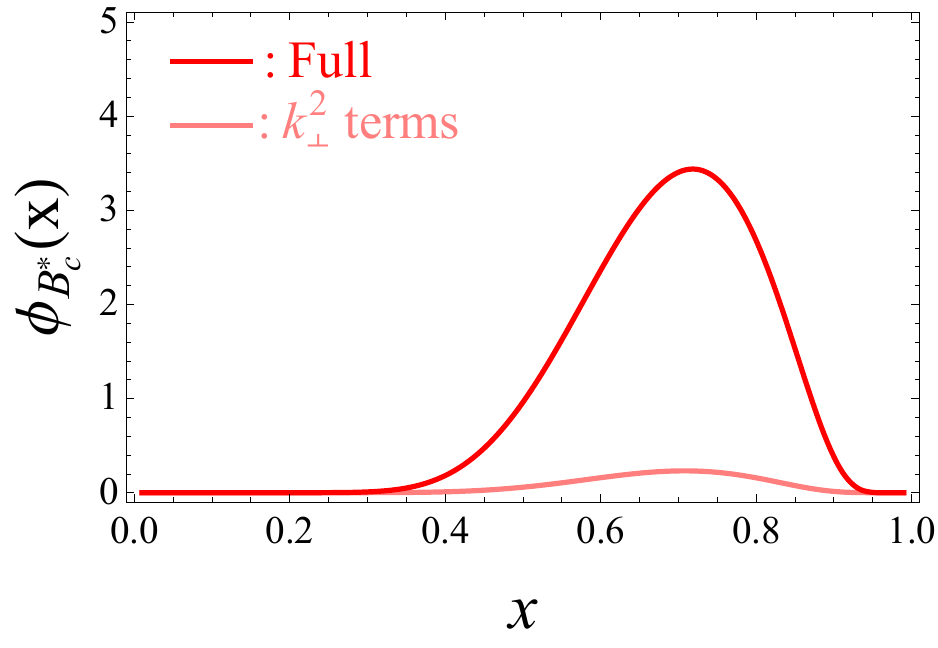}\\
\caption{Full results (red) and $\mathbf{k}_\perp^2$-term contributions (light-colored) for the twist-2 LCDAs of the $K^*$ and $B_c^*$ mesons.}
\label{5}
\end{figure*}

\begin{table*}[!htbp]
\caption{The Gegenbauer moment of twist-2 and twist-3 LCDAs.}
\centering
\begin{tabular*}{\textwidth}{@{\extracolsep{\fill}}c|c c c c |c c c c }
\hline\hline
{meson} & $a_1^\parallel$ & $a_2^\parallel$ & $a_3^\parallel$ & $a_4^\parallel$&  $a_1^\perp$ & $a_2^\perp$& $a_3^\perp$ & $a_4^\perp$ \\
\hline
$\rho$
& 0
& $-0.019^{+0.015}_{-0.017}$
& 0
& $-0.030^{+0.005}_{-0.005}$
& 0
& $0.159^{+0.033}_{-0.034}$
& 0
& $-0.017^{+0.016}_{-0.017}$ \\

$K^*$
& $0.148^{+0.029}_{-0.027}$
& $-0.065^{+0.007}_{-0.006}$
& $0.019^{+0.003}_{-0.003}$
& $-0.026^{+0.004}_{-0.002}$
& $0.220^{+0.036}_{-0.033}$
& $0.086^{+0.010}_{-0.008}$
& $0.077^{+0.002}_{-0.003}$
& $-0.023^{+0.004}_{-0.002}$\\

$D^*$
& $0.633^{+0.055}_{-0.054}$
& $0.126^{+0.051}_{-0.044}$
& $-0.025^{+0.003}_{-0.001}$
& $0.008^{+0.007}_{-0.008}$
& $0.765^{+0.052}_{-0.052}$
& $0.352^{+0.050}_{-0.048}$
& $0.172^{+0.018}_{-0.013}$
& $0.122^{+0.007}_{-0.007}$ \\

$D_s^*$
& $0.517^{+0.055}_{-0.054}$
& $-0.012^{+0.039}_{-0.033}$
& $-0.102^{+0.013}_{-0.009}$
& $-0.035^{+0.010}_{-0.013}$
& $0.608^{+0.055}_{-0.056}$
& $0.124^{+0.044}_{-0.037}$
& $-0.017^{+0.004}_{-0.003}$
& $-0.024^{+0.006}_{-0.004}$ \\

$B^*$
& $1.101^{+0.042}_{-0.044}$
& $0.793^{+0.085}_{-0.085}$
& $0.376^{+0.089}_{-0.082}$
& $0.096^{+0.054}_{-0.042}$
& $1.196^{+0.036}_{-0.038}$
& $1.023^{+0.076}_{-0.077}$
& $0.702^{+0.089}_{-0.085}$
& $0.433^{+0.071}_{-0.061}$\\

$B_s^*$
& $1.031^{+0.027}_{-0.030}$
& $0.645^{+0.053}_{-0.056}$
& $0.202^{+0.052}_{-0.051}$
& $-0.046^{+0.029}_{-0.024}$
& $1.110^{+0.026}_{-0.029}$
& $0.821^{+0.057}_{-0.058}$
& $0.425^{+0.063}_{-0.062}$
& $0.145^{+0.048}_{-0.045}$ \\

$B_c^*$
& $0.621^{+0.063}_{-0.063}$
& $-0.038^{+0.072}_{-0.063}$
& $-0.299^{+0.003}_{-0.004}$
& $-0.165^{+0.048}_{-0.048}$
& $0.662^{+0.063}_{-0.064}$
& $0.023^{+0.075}_{-0.069}$
& $-0.277^{+0.009}_{-0.001}$
& $-0.188^{+0.045}_{-0.042}$ \\
\hline
\end{tabular*}
\label{tab:2}
\end{table*}
\begin{table*}[!htbp]
\caption{The $\xi$-moment of twist-2 and twist-3 LCDAs.}
\centering
\begin{tabular*}{\textwidth}{@{\extracolsep{\fill}}c| c c c c |c c c c }
\hline\hline
{meson} & $\langle \xi^1\rangle^\parallel$ & $\langle \xi^2\rangle^\parallel$& $\langle \xi^3\rangle^\parallel$& $\langle \xi^4\rangle^\parallel$& $\langle \xi^1\rangle^\perp$ & $\langle \xi^2\rangle^\perp$& $\langle \xi^3\rangle^\perp$& $\langle \xi^4\rangle^\perp$\\
\hline
$\rho$
& 0
& $0.193^{+0.006}_{-0.006}$
& 0
& $0.078^{+0.004}_{-0.004}$
& 0
& $0.255^{+0.011}_{-0.012}$
& 0
& $0.120^{+0.009}_{-0.009}$ \\

$K^*$
& $0.089^{+0.017}_{-0.016}$
& $0.178^{+0.002}_{-0.002}$
& $0.042^{+0.007}_{-0.007}$
& $0.068^{+0.001}_{-0.001}$
& $0.132^{+0.022}_{-0.020}$
& $0.229^{+0.004}_{-0.002}$
& $0.071^{+0.010}_{-0.009}$
& $0.103^{+0.002}_{-0.004}$  \\

$D^*$
& $0.380^{+0.033}_{-0.033}$
& $0.243^{+0.018}_{-0.015}$
& $0.158^{+0.015}_{-0.014}$
& $0.115^{+0.011}_{-0.009}$
& $0.459^{+0.031}_{-0.031}$
& $0.321^{+0.018}_{-0.017}$
& $0.229^{+0.017}_{-0.015}$
& $0.179^{+0.013}_{-0.012}$  \\

$D_s^*$
& $0.310^{+0.033}_{-0.032}$
& $0.196^{+0.013}_{-0.012}$
& $0.113^{+0.013}_{-0.011}$
& $0.079^{+0.008}_{-0.006}$
& $0.364^{+0.034}_{-0.033}$
& $0.242^{+0.016}_{-0.012}$
& $0.153^{+0.014}_{-0.014}$
& $0.111^{+0.010}_{-0.008}$  \\

$B^*$
& $0.661^{+0.025}_{-0.027}$
& $0.472^{+0.029}_{-0.029}$
& $0.355^{+0.027}_{-0.027}$
& $0.277^{+0.025}_{-0.024}$
& $0.717^{+0.022}_{-0.022}$
& $0.551^{+0.026}_{-0.027}$
& $0.441^{+0.026}_{-0.026}$
& $0.364^{+0.025}_{-0.024}$  \\

$B_s^*$
& $0.619^{+0.016}_{-0.018}$
& $0.421^{+0.018}_{-0.019}$
& $0.304^{+0.017}_{-0.016}$
& $0.228^{+0.015}_{-0.015}$
& $0.666^{+0.016}_{-0.018}$
& $0.482^{+0.019}_{-0.020}$
& $0.366^{+0.019}_{-0.019}$
& $0.288^{+0.018}_{-0.017}$  \\

$B_c^*$
& $0.373^{+0.037}_{-0.038}$
& $0.187^{+0.025}_{-0.022}$
& $0.102^{+0.017}_{-0.015}$
& $0.060^{+0.011}_{-0.010}$
& $0.397^{+0.038}_{-0.038}$
& $0.207^{+0.027}_{-0.023}$
& $0.117^{+0.019}_{-0.016}$
& $0.071^{+0.011}_{-0.013}$  \\
\hline
\end{tabular*}
\label{tab:3}
\end{table*}
\begin{table*}[!htbp]
\caption{Comparison of $\xi$-moments results in vector mesons with different theoretical models.}
\centering
\begin{tabular*}{\textwidth}{@{\extracolsep{\fill}}c|ccccccc}
\hline
 & This work & LFQM(22)\cite{Arifi:2022pal} & LFQM(19)\cite{Dhiman:2019ddr} & DSE(25)\cite{Xu:2025hjf} & DSE(22)\cite{Serna:2022yfp} & Lattice\cite{Braun:2016wnx,Hua:2020gnw} & Sum Rule\cite{Ball:2007zt} \\
\hline
$\langle \xi^2 \rangle^\parallel_\rho$ & $0.193^{+0.006}_{-0.006}$ & -- & -- & 0.259 & 0.263 & 0.245(9) & 0.234(17) \\
$\langle \xi^2 \rangle^\perp_\rho$ &$0.255^{+0.011}_{-0.012}$  & -- & -- & 0.236 & 0.250 &  0.235(8) &  0.238(17) \\
\hline
$\langle \xi^1 \rangle^\parallel_{K^*}$ & $0.089^{+0.017}_{-0.016}$ & -- & -- & 0.023 & 0.018 &  0.003(4) &  0.0012(12) \\
$\langle \xi^2 \rangle^\parallel_{K^*}$ & $0.178^{+0.002}_{-0.002}$ & -- & -- & 0.220 & 0.272 & 0.205(3)  &  0.227(21) \\
$\langle \xi^1 \rangle^\perp_{K^*}$ & $0.132^{+0.022}_{-0.020}$ & -- & -- & 0.033 & 0.056 &  0.044(4) &  0.0018(18)\\
$\langle \xi^2 \rangle^\perp_{K^*}$ & $0.229^{+0.004}_{-0.002}$ & -- & -- & 0.210 & 0.298 &  0.262(4) &  0.227(21)\\
\hline
$\langle \xi^1 \rangle^\parallel_{D^*}$ & $0.380^{+0.033}_{-0.033}$ & 0.344 & 0.356 & 0.331 & 0.388 & -- & -- \\
$\langle \xi^1 \rangle^\perp_{D^*}$ & $0.459^{+0.031}_{-0.031}$ & -- & 0.351 & 0.356 & 0.484 & -- & -- \\
\hline
$\langle \xi^1 \rangle^\parallel_{D^*_s}$ & $0.310^{+0.033}_{-0.032}$ & 0.296 & 0.323 & 0.308 & 0.254 &--  &--  \\
$\langle \xi^1 \rangle^\perp_{D^*_s}$ & $0.364^{+0.034}_{-0.033}$ & -- & 0.321 & 0.330 & 0.310 & -- &--  \\
\hline
$\langle \xi^1 \rangle^\parallel_{B^*}$ & $0.661^{+0.025}_{-0.027}$ & 0.646 & 0.672 & 0.656 & -- & -- & -- \\
$\langle \xi^1 \rangle^\perp_{B^*}$ & $0.717^{+0.022}_{-0.022}$ & -- & 0.672 & 0.671 & -- & -- &  --\\
\hline
$\langle \xi^1 \rangle^\parallel_{B^*_s}$ & $0.619^{+0.016}_{-0.018}$ & 0.614 & 0.652 & 0.629 & -- & -- & -- \\
$\langle \xi^1 \rangle^\perp_{B^*_s}$ & $0.666^{+0.016}_{-0.018}$ & -- & 0.653 & 0.653 & -- & -- &  --\\
\hline
$\langle \xi^1 \rangle^\parallel_{B^*_c}$ & $0.373^{+0.037}_{-0.038}$ & 0.390 & -- & 0.405 & -- & -- & -- \\
$\langle \xi^1 \rangle^\perp_{B^*_c}$ & $0.397^{+0.038}_{-0.038}$ & -- & -- & 0.416 & -- &  --&  --\\
\hline
\end{tabular*}

\label{tab:4}
\end{table*}
\begin{table*}[!htbp]
\caption{The $\mathbf{k}_\perp$-moment of twist-2 and twist-3 LCDAs.}
\centering
\begin{tabular*}{\textwidth}{@{\extracolsep{\fill}}c| c c c c| c c c c }
\hline\hline
{meson} & $\langle\mathbf{k}_\perp\rangle^\parallel$ & $\sqrt{\langle \mathbf{k}^2_\perp\rangle}^\parallel$& $\sqrt[3]{\langle \mathbf{k}^3_\perp\rangle}^\parallel$& $\sqrt[4]{\langle \mathbf{k}^4_\perp\rangle}^\parallel$ &  $\langle\mathbf{k}_\perp\rangle^\perp$&$\sqrt{\langle \mathbf{k}^2_\perp\rangle}^\perp$& $\sqrt[3]{\langle \mathbf{k}^3_\perp\rangle}^\perp$& $\sqrt[4]{\langle \mathbf{k}^4_\perp\rangle}^\perp$ \\
\hline
$\rho$
& $0.385^{+0.020}_{-0.019}$
& $0.436^{+0.022}_{-0.021}$
& $0.481^{+0.024}_{-0.024}$
& $0.521^{+0.027}_{-0.025}$
& $0.354^{+0.016}_{-0.016}$
& $0.405^{+0.018}_{-0.019}$
& $0.450^{+0.021}_{-0.021}$
& $0.491^{+0.023}_{-0.023}$   \\

$K^*$
& $0.425^{+0.020}_{-0.020}$
& $0.481^{+0.022}_{-0.023}$
& $0.530^{+0.025}_{-0.025}$
& $0.574^{+0.027}_{-0.027}$
& $0.396^{+0.018}_{-0.017}$
& $0.452^{+0.030}_{-0.020}$
& $0.501^{+0.022}_{-0.023}$
& $0.546^{+0.024}_{-0.034}$ \\

$D^*$
& $0.569^{+0.028}_{-0.028}$
& $0.644^{+0.031}_{-0.032}$
& $0.709^{+0.034}_{-0.034}$
& $0.768^{+0.037}_{-0.037}$
& $0.548^{+0.025}_{-0.025}$
& $0.621^{+0.029}_{-0.028}$
& $0.686^{+0.032}_{-0.031}$
& $0.745^{+0.035}_{-0.034}$ \\

$D_s^*$
& $0.637^{+0.031}_{-0.031}$
& $0.720^{+0.035}_{-0.035}$
& $0.793^{+0.039}_{-0.038}$
& $0.859^{+0.042}_{-0.042}$
& $0.615^{+0.028}_{-0.028}$
& $0.697^{+0.033}_{-0.032}$
& $0.770^{+0.035}_{-0.036}$
& $0.835^{+0.039}_{-0.039}$ \\

$B^*$
& $0.658^{+0.032}_{-0.032}$
& $0.744^{+0.036}_{-0.036}$
& $0.820^{+0.040}_{-0.040}$
& $0.888^{+0.044}_{-0.043}$
& $0.646^{+0.030}_{-0.030}$
& $0.732^{+0.034}_{-0.035}$
& $0.807^{+0.038}_{-0.038}$
& $0.875^{+0.042}_{-0.041}$ \\

$B_s^*$
& $0.728^{+0.004}_{-0.003}$
& $0.824^{+0.004}_{-0.004}$
& $0.908^{+0.005}_{-0.005}$
& $0.983^{+0.005}_{-0.005}$
& $0.716^{+0.004}_{-0.003}$
& $0.811^{+0.003}_{-0.003}$
& $0.894^{+0.004}_{-0.003}$
& $0.969^{+0.005}_{-0.005}$ \\

$B_c^*$
& $1.089^{+0.053}_{-0.054}$
& $1.229^{+0.061}_{-0.060}$
& $1.353^{+0.066}_{-0.067}$
& $1.464^{+0.072}_{-0.072}$
& $1.072^{+0.052}_{-0.051}$
& $1.211^{+0.059}_{-0.058}$
& $1.334^{+0.064}_{-0.065}$
& $1.444^{+0.069}_{-0.070}$ \\
\hline
\end{tabular*}
\label{tab:5}
\end{table*}
\subsection{Light-cone Distribution Amplitudes}\label{sec:3.1}

\begin{itemize}
\item As anticipated from SU(2) symmetry, the LCDAs of $\rho$ meson exhibits perfect symmetry about $x=0.5$ as shown in Fig.~\ref{1}. Consequently, all odd-order Gegenbauer moments and $\xi$-moments vanish. There exist substantial differences between the twist-2 and twist-3 LCDAs: the twist-2 LCDA $\phi_{2,\rho}^\parallel(x)$ is narrower, leading to a negative value of the Gegenbauer moment $a_2^\parallel$, whereas the twist-3 LCDA $\phi_{3,\rho}^\perp(x)$ is flatter, resulting in a positive value of $a_2^\perp$, as summarized in Table~\ref{tab:2}.

\item The right panel of Fig.~\ref{1} shows the LCDAs of the $K^*$ meson. Due to SU(3) flavor symmetry breaking, both $\phi_{2,K^*}^\parallel(x)$ and $\phi_{3,K^*}^\perp(x)$ show noticeable asymmetry. This indicates that the heavier constituent quark carries a larger fraction of the longitudinal momentum than the lighter one, implying nonzero $\xi$-moments. The symmetry breaking effect in $\phi_{3,K^*}^\perp(x)$ is more pronounced than that in $\phi_{2,K^*}^\parallel(x)$, consistent with observations in studies of pseudoscalar LCDAs~\cite{Li:2026wmb}, where symmetry breaking becomes increasingly significant for higher-twist LCDAs. This feature will be further confirmed by the numerical analysis of the $\xi$-moments $\langle \xi^n\rangle^{\parallel(\perp)}$. Since higher-twist LCDAs encode contributions from the transverse motion of the constituents, this behavior suggests that transverse dynamics are particularly sensitive to the quark mass difference.

\item The twist-2 and twist-3 LCDAs of the pion and kaon are also presented in Fig.~\ref{2}, together with those of $\rho$ and $K^*$ for comparison. With the replacement $M\to M_0$, the shapes of the curves for $\phi^{A}_{2,\pi}(x)$ and $\phi^P_{3,\pi}(x)$ become identical, as evidenced by the completely overlap of the blue and red dashed lines in left panel of Fig.~\ref{2}. Nevertheless, significant differences remain between the LCDAs of the $\pi$ and $\rho$ mesons. For the $K$ and $K^*$ mesons, the curves of $\phi_{2}(x)$ are close to each other, as are those of $\phi_{3}(x)$, suggesting a similarity between the LCDAs of pseudoscalar and vector mesons for the same twist order.
\end{itemize}

\begin{itemize}
\item The results of the heavy vector mesons are shown in Fig.~\ref{3}, where both twist-2 and twist-3 LCDAs exhibit pronounced endpoint behavior: substantial contributions appear near $x \approx 1$, corresponding to large longitudinal momentum fractions carried by the heavy quark, while the contributions are nearly vanishing near $x \approx 0$, where the light quark carries only a small fraction of the longitudinal momentum. Notably, in the $B_c^*$ case, the LCDAs tend to zero at both endpoints, similar to  the LCDAs of $B_c$~\cite{Li:2026wmb}. Comparing the twist-2 and twist-3 LCDAs curves, the flavor symmetry breaking effects are more pronounced in the higher twist LCDAs, extending the observations discussed above. Interestingly, a clear trend emerges: as the constituent quark mass (or meson mass) increases, the differences between LCDAs of different twists gradually diminish; for instance, the two LCDAs of the $B_c^*$ meson are already very close. This behavior is consistent with the studies of quarkonium LCDAs~\cite{Xu:2026zli}, where it was found that $\phi_{2}^\parallel(x)$ and $\phi_{3}^\perp(x)$ become identical in heavy quark limit, i.e., $\phi_{2}^\parallel(x)\approx\phi_{3}^\perp(x)$ as $m_q\rightarrow\infty$, indicating a weak dependence of the LCDAs on the twists in heavy quark limit. It should be noted that this behavior could be understood as a model-dependent feature within the self-consistent LFQM, rather than a general prediction of QCD. More precisely, the observed suppression of twist dependence in the heavy-quark limit should be interpreted as a robust qualitative trend within the self-consistent LFQM framework, rather than a strict model-independent consequence derived directly from QCD.

\item Fig.~\ref{4} also displays the LCDAs of the corresponding pseudoscalar mesons, marked by dashed lines. A comparison reveals that the differences between the twist-2 curves of $\phi^A_2(x)$ and $\phi_2^\parallel(x)$ decrease with increasing meson mass; for example, the curves of $B_c$ and $B_c^*$ almost coincide. It can therefore be anticipated that, in the heavy quark limit $m\rightarrow\infty$, they are consistent with each other, completely in agreement with the studies on quarkonium~\cite{Xu:2026zli}. Further calculations show that the integral contribution from the $\mathbf{k}^2_\perp$ terms in the Eqs.~(\ref{eq:13}) and (\ref{eq:14}) are not dominant, accounting for about $30\%$ for light mesons and less than $10\%$ for heavy mesons, as illustrated in Fig.~\ref{5}. For clarity, we present here the explicit expression of $\phi_2^A$ given in Ref.~\cite{Li:2026wmb} for comparison:
     \begin{align}
     &&\phi^A_2(x,\mu) = \frac{\sqrt{2N_c}}{f_M}\int_0^{\mu^2} \frac{d^2\mathbf{k}_\perp}{8\pi^3}\frac{\psi(x,{\bf{k}}_{\perp})(\bar{x}m_q+xm_{\bar{q}})}{\sqrt{(\bar{x}m_q+xm_{\bar{q}})^2
     +\mathbf{k}^2_\perp}}.\nonumber
     \end{align}
Consequently, in the heavy-quark limit, the relation
$\phi_2^A(x)\simeq\phi_2^\parallel(x)$ holds, indicating a weak spin dependence of the LCDAs within the self-consistent LFQM framework. For the same reason, one also finds $\phi_3^P(x)\simeq\phi_3^\perp(x)$. Combining this behavior with the suppression of twist dependence discussed above, one obtains the approximate relation
$\phi_2^A(x) \simeq \phi_2^{\parallel}(x) \approx\phi_3^P(x) \simeq \phi_3^{\perp}(x)$
in the heavy-quark limit. We emphasize that this feature is closely related to the LFQM and is not expected to represent a universal model-independent prediction of QCD. It is also found that this behavior remains stable for reasonable ranges of the Gaussian parameter $\beta$, while the dependence on the radial wave function is worth to be further explored. Nevertheless, the observed suppression of spin-dependent differences is qualitatively consistent with the general expectation of heavy-quark spin symmetry in HQET/NRQCD, where such effects are suppressed by ${\cal O}(\Lambda_{\rm QCD}/m_Q)$ corrections.
\end{itemize}

\subsection{Gegenbauer moments, $\xi$-moments and transverse momentum moments}\label{sec:3.2}
The $n$th Gegenbauer moment $a_n(\mu)$, $\xi$-moment $\langle \xi^n\rangle^{\parallel(\perp)}$, and transverse moment $\langle \mathbf{k}^n_\perp\rangle^{\parallel(\perp)}$ are calculated, and the numerical results are presented in Tables.~\ref{tab:2}--\ref{tab:4}. From these results, one can make the following observations:

\begin{itemize}
\item As mentioned above, due to SU(2) symmetry, the odd-order Gegenbauer moments $a_{1,3}$ of $\rho$ vanish. In general, the value of $a_1$ increases with the heavy quark mass and decreases with the reduction in the mass difference between constituent quarks. The values of $a_{1}$ for the twist-3 are larger than those of twist-2, i.e., $a_1^\perp>a_1^\parallel$, as shown in Table.~\ref{tab:2}, which is similar to the findings in study of pseudoscalar mesons~\cite{Li:2026wmb}. This also confirms that the symmetry breaking effect in higher twist DAs is more remarkable. Similar to the Gegenbauer moments, the values of $\xi$-moments generally increases with the heavy quark mass and decreases with the reduction in the mass difference between constituent quarks. The values of $\langle \xi^n\rangle$ for the twist-3 are generally larger than those of twist-2, i.e., $\langle \xi^n\rangle^\perp>\langle \xi^n\rangle^\parallel$, as shown in Table.~\ref{tab:3}. This reflects a greater longitudinal separation between the constituent quarks in the higher twist LCDAs. It is also evident that $\langle \xi^n\rangle$ naturally decrease as $n$ increases. For brevity, we compare some of our results with other available theoretical models, as summarized in Table.~\ref{tab:4}. One noticeable discrepancy appears in the values of $\langle \xi^1\rangle_{K^*}^{\parallel,\perp}$, which are larger than those obtained in DSE, lattice QCD, and QCD sum-rule analyses. This difference mainly originates from the stronger SU(3) flavor-symmetry breaking effects in the present LFQM framework, where the constituent strange-quark mass and the self-consistent $M\rightarrow M_0$ treatment enhance the asymmetry of the longitudinal momentum distribution. In addition, different nonperturbative approaches adopt distinct treatments of confinement dynamics, quark masses, and wave-function structures, which can further lead to quantitative differences in the $\xi$-moments. Nevertheless, the overall hierarchical behavior and qualitative trends remain consistent among different approaches.

\item The transverse moment $\langle \mathbf{k}^n_\perp\rangle^{\parallel(\perp)}$ generally increases with the parameter $\beta$ (and thus with meson mass), supporting the interpretation that $\beta$ characterizes the transverse ``size'' of the bound state and implying that transverse separation of constituent quarks is more pronounced in heavy mesons. Interestingly, the twist-2 LCDAs exhibit slightly larger transverse moments than the twist-3 LCDAs, i.e., $\langle \mathbf{k}_\perp^n\rangle^{\parallel} \gtrsim \langle \mathbf{k}_\perp^n\rangle^{\perp}$. Table.~\ref{tab:5} further shows that for the heavy mesons, the difference between the twist-2 and twist-3 transverse moment are very small (less than $2\%$). Therefore, one can expects that $\langle \mathbf{k}^n_\perp\rangle^{\parallel(\perp)}$ will exhibit twist independence in the heavy quark limit, which consist with the findings in the previous studies of pseudoscalar mesons showing that transverse moments $\langle \mathbf{k}^n_\perp\rangle$ are twist independent under replacement $M\to M_0$~\cite{Li:2026wmb}.
\end{itemize}

\section{Summary}\label{sec:4}

In this work, we investigate the twist-2 and twist-3 LCDAs of vector mesons within the LFQM. The numerical analysis is performed using parameters fixed by constraints from mesonic decay constants. In addition, the Gegenbauer moments $a_n$, the $\xi$-moments $\langle \xi^n\rangle$, and the transverse moments $\langle \mathbf{k}_\perp^n\rangle$ have also been analyzed. Our main findings are as follows:

(i) Flavor symmetry breaking effects are more pronounced in the twist-3 LCDAs of vector mesons, leading to $a_1^\perp > a_1^\parallel$ and $\langle \xi^n\rangle^\perp > \langle \xi^n\rangle^\parallel$. This is consistent with previous results for the LCDAs of pseudoscalar mesons.

(ii) For vector mesons, the twist dependence is suppressed in the heavy-quark limit, leading to $\phi_{2}^\parallel(x)\simeq\phi_{3}^\perp(x)$. Moreover, for pseudoscalar and vector mesons composed of the same quark constituents, LCDAs of the same twist become increasingly similar and gradually converge as the quark mass increases; in the limit $m_q\to\infty$, one finds $\phi_{2}^A(x)\simeq\phi_{2}^\parallel(x)$ and $\phi_{3}^P(x)\simeq\phi_{3}^\perp(x)$, indicating the spin independence of LCDAs. Within the self-consistent LFQM framework, the LCDAs satisfy $\phi_2^A(x) \simeq \phi_2^{\parallel}(x) \approx\phi_3^P(x) \simeq \phi_3^{\perp}(x)$
in the heavy-quark limit. This behavior should be interpreted as a model-dependent realization of the suppression of spin-dependent effects, qualitatively consistent with heavy-quark spin symmetry expectations, which provides a new perspective on the relation between pseudoscalar and vector meson structures within the LFQM framework.

\section*{ACKNOWLEDGMENTS}
We would like to thank Prof.Fu-Sheng Yu for valuable inspiration. Xiao-Nan Li is supported by the Anhui Provincial Department of Education Scientific Research Project (Grant No.2025AHGXZK40010). Qin Chang is supported by the National Natural Science Foundation of China (Grant No.12275067), Science and Technology R$\&$D Program Joint Fund Project of Henan Province (Grant No.225200810030), Science and Technology Innovation Leading Talent Support Program of Henan Province (Grant No.254200510039), and National Key R$\&$D Program of China (Grant No.2023YFA1606000).

\end{document}